\documentclass[fleqn,twoside]{article}
\usepackage{espcrc2}

\usepackage{graphicx}
\usepackage{epsfig}
\usepackage[figuresright]{rotating}

\newcommand{\postscript}[2]{\setlength{\epsfxsize}{#2\hsize}
   \centerline{\epsfbox{#1}}}

\newcommand{\gev}{\text{GeV}}
\newcommand{\tev}{\text{TeV}}

\newcommand{\mgaugino}{M_{1/2}}
\newcommand{\mweak}{M_{\text{weak}}}
\newcommand{\mplanck}{M_{\text{Planck}}}

\newcommand{\mgut}{M_{\text{GUT}}}

\newcommand{\OmegaDM}{\Omega_{\text{DM}}}

\newcommand{\eqref}[1]{Eq.~(\ref{#1})}

\newcommand{\figref}[1]{Fig.~\ref{fig:#1}}

\newcommand{\text}[1]{{\rm #1}}

\newcommand{\agt}{ \mathop{}_{\textstyle \sim}^{\textstyle >} }
\newcommand{\alt}{ \mathop{}_{\textstyle \sim}^{\textstyle <} }

\title{Dark Matter Detection in Space}

\author{Jonathan~L.~Feng\address{Department of Physics and Astronomy,
University of California, Irvine, CA 92697}}
      
\begin{document}

\begin{abstract}
I review prospects for detecting dark matter in space-based
experiments, with an emphasis on recent developments.  I propose the
``Martha Stewart criterion'' for identifying dark matter candidates
that are particularly worth investigation and focus on three that
satisfy it: neutralino dark matter, Kaluza-Klein dark matter, and
superWIMP gravitino dark matter.
\vspace{1pc}
\end{abstract}

\maketitle

\section{Dark Matter Now}

We live in interesting times: we know dark matter exists, and we now
know how much.  At the same time, we have very little idea what it is,
other than that it can't be any known particle.  Cosmology therefore
provides the long-awaited quantitative evidence for new particle
physics.  The dark matter problem is arguably our strongest beacon, an
unambiguous and fundamental, yet approachable, problem pointing us
toward promising directions at the frontier.

As with most frontiers, the dark matter landscape is populated by a
variety of colorful characters:
axions~\cite{Peccei:1977ur,Wilczek:pj,Weinberg:1977ma},
neutralinos~\cite{Goldberg:1983nd,Ellis:1983ew}, Q
balls~\cite{Kusenko:1997si}, wimpzillas~\cite{Chung:1998ua},
axinos~\cite{Covi:1999ty}, self-interacting dark
matter~\cite{Spergel:1999mh}, fuzzy dark matter~\cite{Hu:ti},
annihilating dark matter~\cite{Kaplinghat:2000vt}, Kaluza-Klein dark
matter~\cite{Servant:2002aq,Cheng:2002ej}, superWIMP dark
matter~\cite{Feng:2003xh,Feng:2003uy}, and many others.  To bring some
order to the following discussion, if not to the field, we therefore
need a guiding principle.

\section{The Martha Stewart Criterion}

The naturalness with which the observed relic density is obtained
provides a selection criterion.  Any proposal for dark matter must
explain the observed relic density, the one piece of rather precise
quantitative information we have about dark matter.  For many dark
matter candidates, the relic density may vary over many orders of
magnitude, and the observed relic density is obtained by fine-tuning
one or more free parameters. On the other hand, for some dark matter
candidates the relic density is automatically in the correct range
without the need to introduce and adjust new mass scales.

\begin{figure*}[tbh]
\vspace{9pt}
\begin{minipage}[t]{0.44\textwidth}
\begin{center}
\postscript{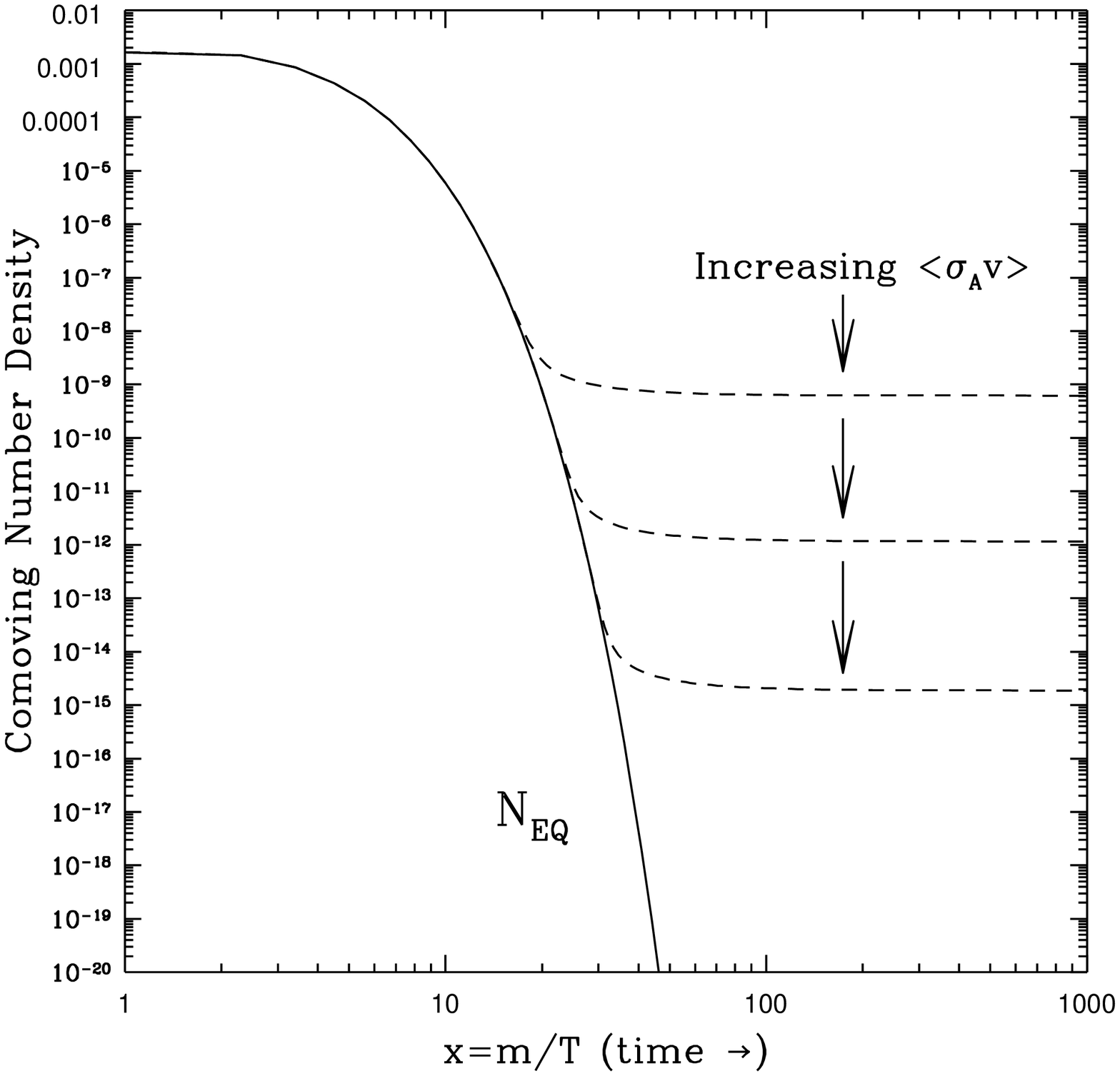}{0.87}
\end{center}
\end{minipage}
\hfil
\begin{minipage}[t]{0.54\textwidth}
\begin{center}
\postscript{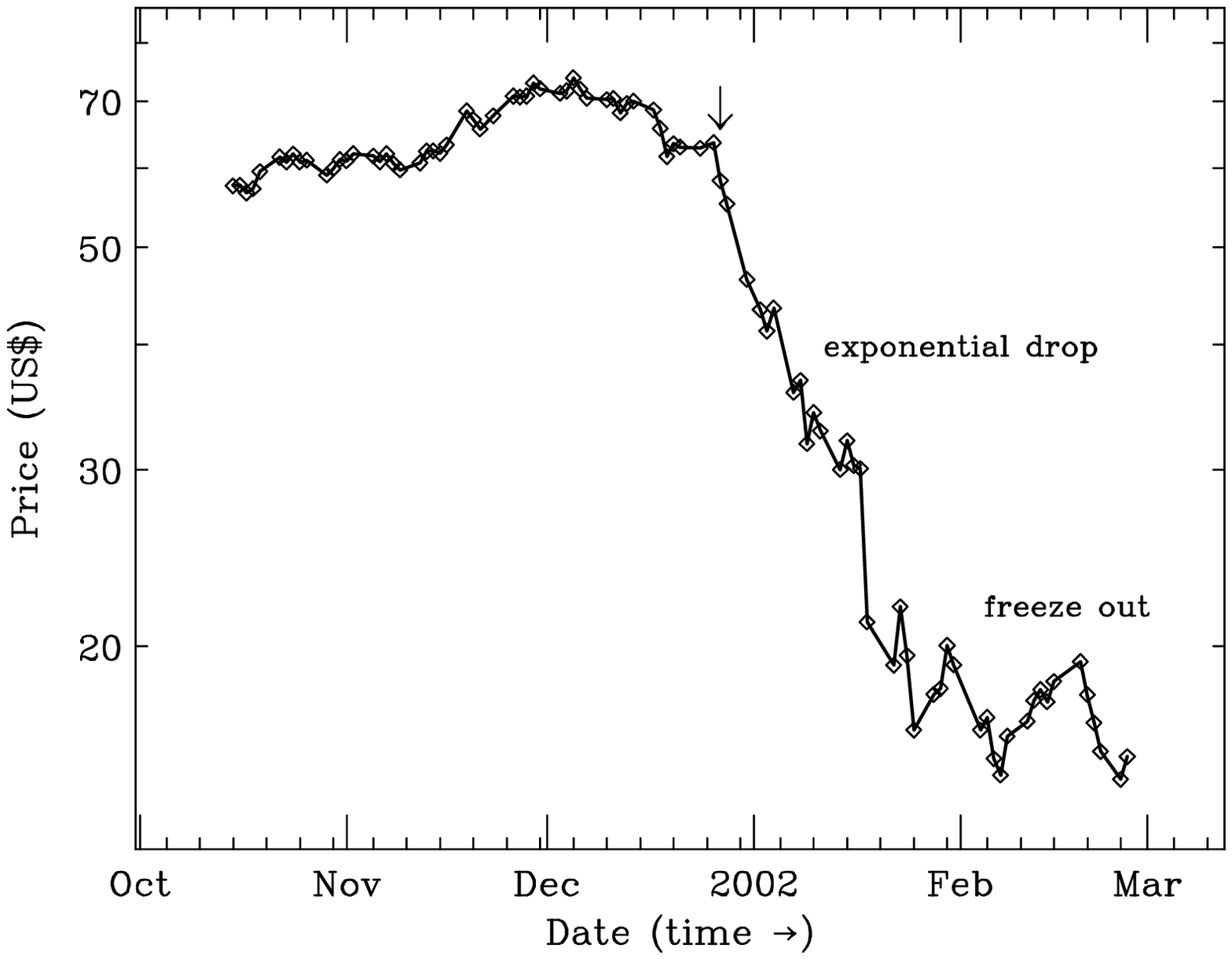}{0.87}
\end{center}
\end{minipage}
\vspace*{-.42in}
\caption{The number density of massive stable particles (left) and the
price of ImClone stock (right) as functions of time. In the right
panel, the date of Martha Stewart's stock sale is indicated by the
arrow.}
\label{fig:freezeout}
\end{figure*}

In the latter category, the most well-known examples are
weakly-interacting massive particles (WIMPs).  WIMP masses and
annihilation cross sections are set by the weak scale, as dictated by
particle physics considerations alone.  In the early universe, all
particles are in thermal equilibrium.  As the universe cools, the
number of stable, massive particles falls, eventually dropping
exponentially with the Boltzmann suppression factor $e^{-m/T}$.  Soon
after that, however, when the annihilation rate falls below the
expansion rate, the number of particles approaches a constant --- the
particles ``freeze out.''  This behavior is shown in
\figref{freezeout}.  It is well-known that for WIMPs, the freeze out
relic density, a function of only the known energy scales $\mweak \sim
100~\gev$ and $\mplanck \sim 10^{19}~\gev$, is naturally near the
observed value $\Omega_{\text{DM}} \sim 0.1$.

What are we to make of this fact?  Consider an analogous phenomenon
that occurred much later in the history of the universe.  On 27
December 2001, the American lifestyle guru Martha Stewart sold all of
her ImClone stock.  The next day, the price of ImClone stock began an
exponential drop, eventually freezing out at a price roughly 1/4 its
initial value.  The ImClone stock price as a function of time is also
shown in \figref{freezeout}.  

Is dark matter made of WIMPs?  Is Ms.~Stewart guilty?  Both cases
initially rest on what may simply be coincidences.  However, in both
cases, these coincidences are so remarkable that they warrant serious
investigation.  Here we will discuss three dark matter candidates that
satisfy the ``Martha Stewart criterion'': they naturally obtain the
observed relic density without the need to introduce and fine-tune new
energy scales.  There are, of course, other highly-motivated
candidates, such as axions.  Note, however, that an additional benefit
of studying candidates that pass the Martha Stewart criterion is that
the absence of completely unknown energy scales makes these scenarios
relatively predictive. They are typically open to investigation from
many directions, and so may be explored by a rich variety of methods
at the interface of particle physics, astrophysics, and cosmology.

\section{Neutralinos}

Neutralinos $\chi$ are the prototypical WIMPs.  Neutralinos are in
general mixtures of the superpartners of neutral Higgs and gauge
bosons.  Particle physics considerations require their mass and
interactions to be determined by the weak scale, and so they naturally
freeze out with relic density near $\Omega_{\text{DM}} \sim 0.1$, as
discussed above.  Note, however, that although neutralinos have become
leading dark matter candidates, they are prototypical, but not
typical.  Not only are they fermions, which is not a WIMP requirement,
but they are Majorana fermions (they are their own anti-particles).
The latter property has strong consequences, as we will see.

Neutralinos may be detected directly by looking for scattering in
highly sensitive detectors below the Earth's surface.  The prospects
for direct detection have been reviewed by Konstantin
Matchev~\cite{Matchev:2004pr}.  Here our focus is on space-based
detection.  We therefore consider indirect detection in which dark
matter particles annihilate with each other somewhere in the universe,
and their annihilation products are detected in space.

A leading indirect signal is positrons from $\chi \chi$ annihilation
in the galactic halo~\cite{Turner:1990kg,Kamionkowski:1991ty}.  Hard
positrons are the best signal, as the background drops rapidly as
positron energy increases, and the background is also better
understood at high energies.  The best possible signal, then, would be
from $\chi \chi \to e^+ e^-$.  Unfortunately, because neutralinos are
Majorana fermions, the Pauli exclusion principle implies that a pair
of neutralinos in an initial $S$-wave state has total angular momentum
$J = 0$.  This process is therefore extremely suppressed, either by
the $P$-wave factor $v^2 \sim 10^{-6}$, where $v$ is the average WIMP
velocity now, or by $m_e / \mweak \sim 10^{-5}$.

The next best hope for hard positrons if from $\chi \chi \to W^+ W^-,
ZZ$, followed by gauge boson decay to positrons.  In many models,
however, the neutralino is Bino-like, that is, it is the superpartner
of the U(1) hypercharge gauge boson.  In this case, it does not couple
to SU(2) gauge bosons, and these modes are also suppressed.  For
Bino-like neutralinos, then, the leading sources of positrons are
processes such as $\chi \chi \to \bar{b} b$, followed by $\bar{b} \to
\bar{c} e^+ \nu$.  These signals are far from ideal, as the 3-body
decays produce broad and soft positron energy distributions.

Another prominent indirect signal is photons from neutralino
annihilation in the center of our
galaxy~\cite{Berezinsky:1992mx,Urban:1992ej}.  The best signal is hard
photons, but again the best hope, $\chi \chi \to \gamma \gamma$, is
highly suppressed, as it is possible only through loop diagrams.  The
next best signal is $\chi \chi \to W^+ W^- , ZZ$ followed by gauge
bosons decaying eventually to photons.  As noted above, however, for
Bino-like neutralinos, this is absent, and one must turn to $\chi \chi
\to f \bar{f}$, resulting in relatively soft photons.

These points are nicely illustrated in minimal supergravity, a simple
model that incorporates many of the virtues of weak-scale
supersymmetry.  Minimal supergravity is parametrized by 5 parameters,
the most important of which are $m_0$ and $\mgaugino$, the unified
scalar and gaugino masses at the grand unified scale $\mgut \simeq 2
\times 10^{16}~\gev$.

\begin{figure*}[tbh]
\vspace{9pt}
\begin{minipage}[t]{0.46\textwidth}
\begin{center}
\postscript{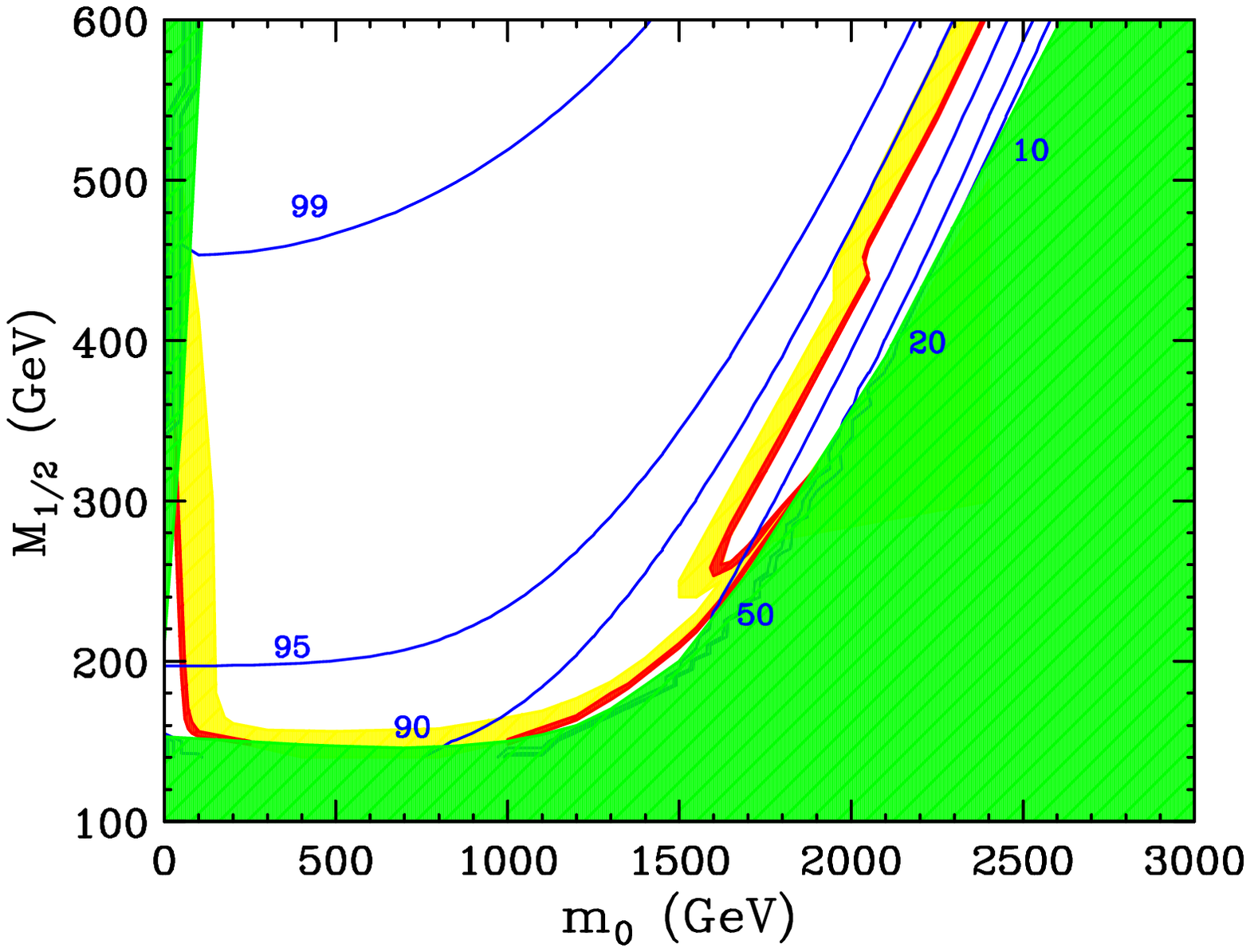}{0.97}
\end{center}
\end{minipage}
\hfil
\begin{minipage}[t]{0.52\textwidth}
\begin{center}
\postscript{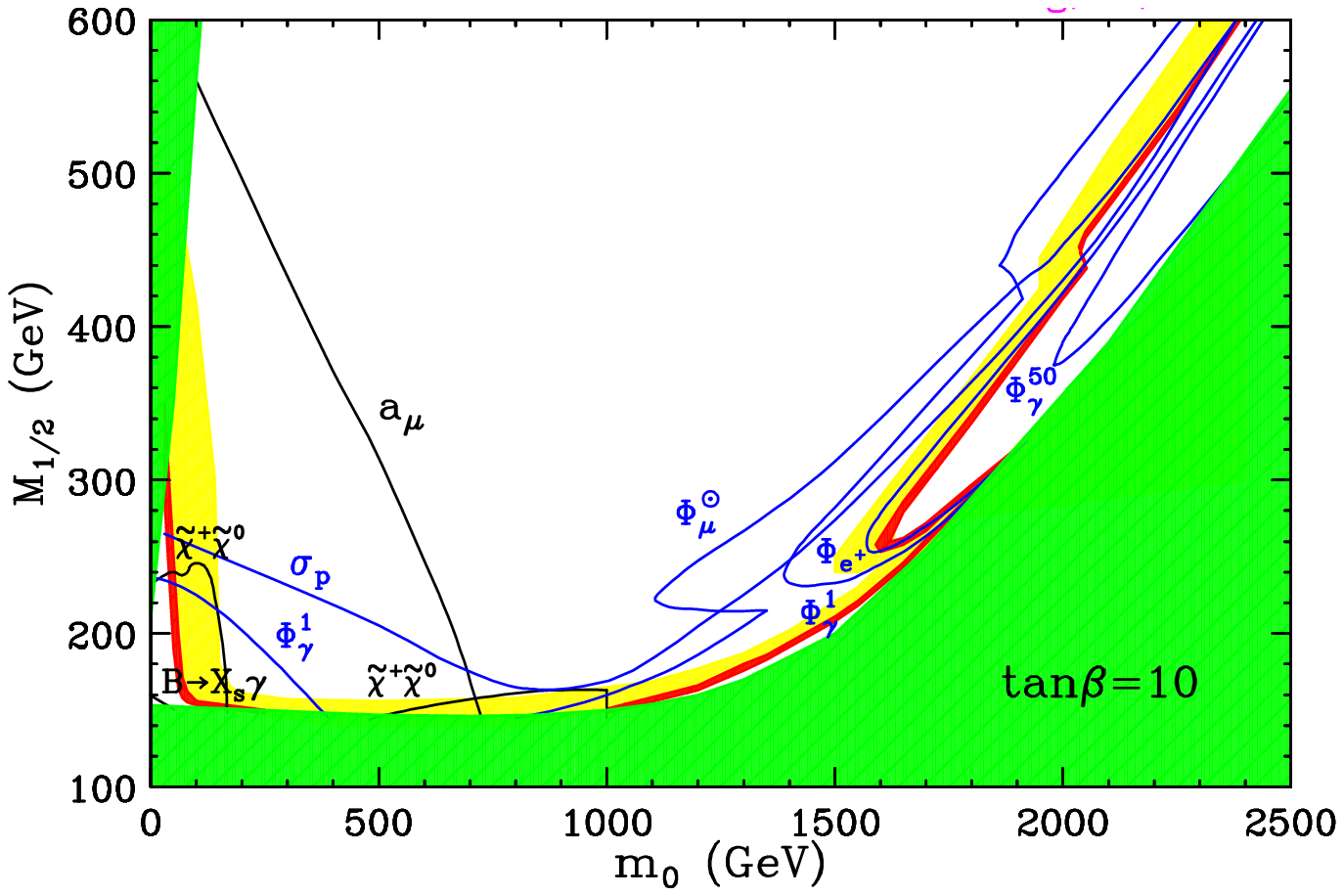}{0.97}
\end{center}
\end{minipage}
\vspace*{-.42in}
\caption{Left: Contours of neutralino gaugino-ness $R_{\chi} \equiv
  |a_{\tilde{B}}|^2 + |a_{\tilde{W}}|^2$ in percent, where $\chi =
  a_{\tilde{B}} (- i \tilde{B}) + a_{\tilde{W}} (- i \tilde{W}) +
  a_{\tilde{H}_d} \tilde{H}_d + a_{\tilde{H}_u} \tilde{H}_u$, in the
  $(m_0, \mgaugino)$ plane of minimal supergravity with $A_0 = 0$,
  $\tan\beta=10$, and $\mu >0$~\protect\cite{Feng:2000gh}.  The green
  (medium shaded) regions are excluded.  In the yellow (light shaded)
  region, the thermal relic density satisfies the pre-WMAP constraint
  $0.1 < \OmegaDM h^2 < 0.3$. In the red (dark shaded) region, the
  neutralino density is in the post-WMAP range $0.094 < \OmegaDM h^2 <
  0.129$.  Right: Reaches of various high-energy collider and
  low-energy precision searches (black) and direct and indirect dark
  matter searches (blue) in the next few
  years~\protect\cite{Feng:2000zu}.  The experiments probe regions
  below the contours indicated, and the shaded regions are as in the
  left panel.}
\label{fig:msugra}
\end{figure*}

A slice of minimal supergravity parameter space is shown in
\figref{msugra}.  In each panel, the region with the currently favored
range of dark matter density~\cite{Spergel:2003cb,Tegmark:2003ud} is
red (dark shaded).  This favored region has two branches.  The region
with $m_0 \alt 200~\gev$ is known as the ``bulk region.''  Here the
neutralino is Bino-like and space-based indirect searches are
therefore weak.  However, in the cosmologically preferred region with
$m_0 \agt 1~\tev$, the ``focus point
region''~\cite{Feng:1999mn,Feng:1999zg}, the neutralino is a
Bino-Higgsino mixture, and space-based indirect searches are quite
promising.  Particularly relevant for the present discussion are the
contours labeled $\Phi_{e^+}$ and $\Phi_{\gamma}^1$, which denote the
reaches of AMS and GLAST, respectively.

Given a specific model, such as minimal supergravity, the properties
of many supersymmetric particles are correlated, and a broad range of
data may have implications for dark matter.  In fact, recent progress
in particle physics and cosmology now disfavors the bulk region.  In
particular, the combination of the Higgs boson mass bound $m_h >
115~\gev$, the consistency of $B(b \to s \gamma)$ (and, possibly,
$(g-2)_{\mu}$) with standard model predictions, and the low value of
$\Omega_{\text{DM}}$ favored by WMAP essentially excludes the bulk
region.  The focus point region remains as one of the viable
alternatives, however.  Recent progress therefore enhances the
possibility that supersymmetric dark matter may have properties
accessible to indirect detection through space-based experiments.
Although this discussion has been confined to minimal supergravity,
the line of argument is valid more generally and applies to model
frameworks beyond minimal supergravity.

\section{Kaluza-Klein Dark Matter}

Another recent development is the investigation of dark matter
candidates in models with extra spatial dimensions.  Such models
generically predict a tower of Kaluza-Klein (KK) particles for every
field that propagates in the extra dimensions.  It is natural to
attempt to identify one of these KK particles with dark matter.

Of particular interest are models with universal extra dimensions, in
which all standard model fields propagate~\cite{Appelquist:2000nn}.
If the extra dimensions are compactified on circles, such models
predict massless states that have not been observed.  To eliminate
such states, one may compactify the extra dimensions on orbifolds.
Such compactifications not only eliminate the unwanted states, but
they also preserve a discrete symmetry, called KK-parity, which
ensures the stability of the lightest KK particle.  Particle physics
provides motivations for tying the KK particle masses to the weak
scale, and for expecting the lightest KK particle to be neutral in
charge and color~\cite{KKrefs}.  In such models, then, an excellent
dark matter candidate naturally emerges --- a stable WIMP with relic
density naturally in the right range.

Dark matter in universal extra dimension models is similar to dark
matter in supersymmetry, with one key difference.  In supersymmetry,
superpartners differ in spin by 1/2, but in extra dimensions, the
excited KK states have the same spin as their ground state standard
model partners.  This has profound implications for dark matter.
Suppose the lightest KK particle is the $B^1$ boson, the first excited
KK state of the hypercharge gauge boson.  Unlike the Bino, the $B^1$
is a vector particle with spin 1.  The suppressions applicable to
Majorana fermion dark matter therefore do not apply.  For example, the
process $B^1 B^1 \to e^+ e^-$ is unsuppressed, and, in fact, $\sim
20\%$ of $B^1$ annihilations occur through this channel.

\begin{figure*}[tbh]
\vspace{9pt}
\begin{minipage}[t]{0.49\textwidth}
\begin{center}
\postscript{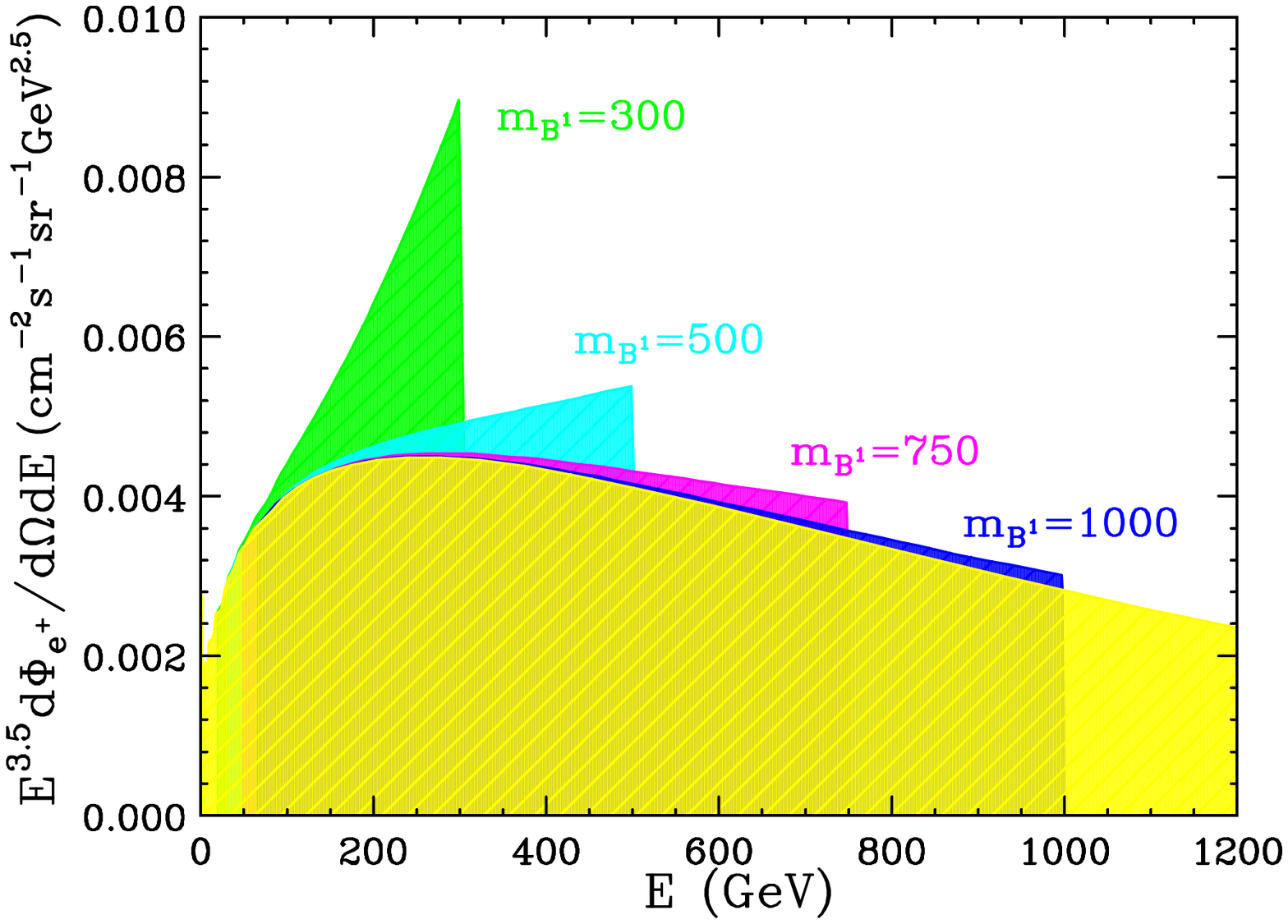}{0.97}
\end{center}
\end{minipage}
\hfil
\begin{minipage}[t]{0.49\textwidth}
\begin{center}
\postscript{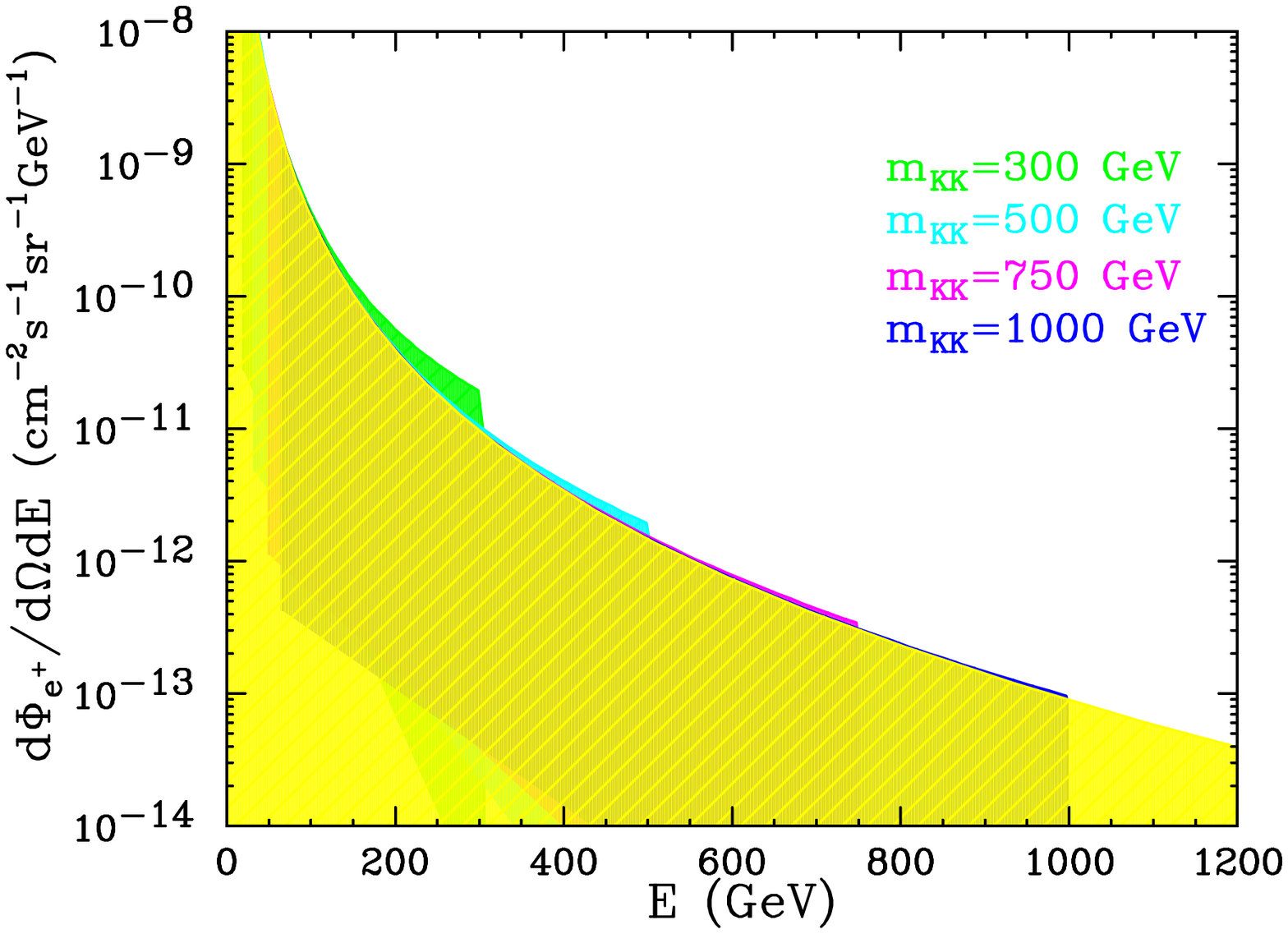}{0.97}
\end{center}
\end{minipage}
\vspace*{-.42in}
\caption{Positron spectra from $B^1$ dark matter annihilation for
  various $B^1$ masses as indicated~\protect\cite{Feng:2000zu}.  The
  yellow (light shaded) region is the expected background.  The
  differential flux is given in the right panel, and is modified by
  the factor $E^3$ in the left panel.}
\label{fig:KKDM}
\end{figure*}

Because the $B^1$ dark matter is highly non-relativistic now,
positrons are therefore produced mono-energetically with energy equal
to the dark matter mass.  Positron spectra for various $B^1$ dark
matter masses are shown in \figref{KKDM}.  The mono-energetic spike is
modified by propagation in the galactic halo~\cite{Moskalenko:1999sb}.
Despite this, we see that the positron signal retains a characteristic
sharp upper edge.  Detection of such a feature by the space-based
experiments PAMELA and/or AMS would signal the discovery dark matter.
At the same time, it would simultaneously exclude the soft and broad
spectra predicted by neutralino dark matter, and would even provide a
measurement of the dark matter particle's mass.

\section{SuperWIMP Gravitino Dark Matter}

Both supersymmetry and extra dimensions predict partner particles for
all known particles.  Can a partner of the graviton, either the
gravitino or a KK graviton, be the dark matter? Such particles are
interesting, as they are minimal dark matter candidates in that they
interact only through gravity, the only interaction required of dark
matter.

Let us consider the gravitino.  We assume it is the lightest
supersymmetric particle~\cite{Pagels:ke,Weinberg:zq}.  Given its
extremely weak couplings, it will play no role in the early
universe.\footnote{We assume that primordial number densities are
diluted by inflation, and the universe reheats to a low enough
temperature that there is no significant regeneration.  For
gravitinos, this is valid for reheat temperatures $T_{\text{RH}} \alt
10^8 - 10^{10}~\gev$~\cite{Bolz:2000fu}; for KK gravitons, it requires
$T_{\text{RH}} \alt 1 - 100~\tev$~\cite{Feng:2003nr}.}  If the
next-lightest supersymmetric particle is a WIMP, it will freeze out
with a thermal relic density in the preferred range.  However, it
eventually decays to the gravitino with lifetime
\begin{equation}
\tau = \Gamma^{-1} \sim \mplanck^2/\mweak^3 \sim \ \text{year} \ .
\end{equation}

The gravitino then inherits the desired relic density.  As with
conventional WIMPs, the gravitino relic density is determined by
$\mweak$ and $\mplanck$ only.  This scenario therefore satisfies the
Martha Stewart criterion.  Note, however, that the gravitino is not a
WIMP; rather it is a {\em super}weakly-interacting massive particle,
or superWIMP.

Decays to gravitino superWIMPs occur long after Big Bang
nucleosynthesis (BBN).  One might therefore guess that the superWIMP
scenario is excluded if one wants to preserve the successful BBN
predictions for the light element abundances~\cite{Ellis:1990nb}. The
dominant form of visible energy released in decays to superWIMPs is
electromagnetic.  For such energy, BBN constraints are conveniently
given in the $(\tau_{\text{WIMP}}, \zeta_{\text{EM}})$ plane.  Here
$\tau_{\text{WIMP}}$ is the WIMP lifetime, the time at which the
energy is released, and
\begin{equation}
\zeta_{\text{EM}} \equiv \epsilon_{\text{EM}} Y_{\text{WIMP}}
\label{zeta}
\end{equation}
is a measure of the energy released.  In \eqref{zeta},
$\epsilon_{\text{EM}}$ is the initial electromagnetic energy released
in each WIMP decay, and $Y_{\text{WIMP}} \equiv
n_{\text{WIMP}}/n_{\gamma}^{\text{BG}}$ is the WIMP number density
before they decay, normalized to the number density of background
photons $n_{\gamma}^{\text{BG}} = 2 \zeta(3) T^3/\pi^2$.

\begin{figure*}[tbh]
\vspace{9pt}
\begin{center}
\postscript{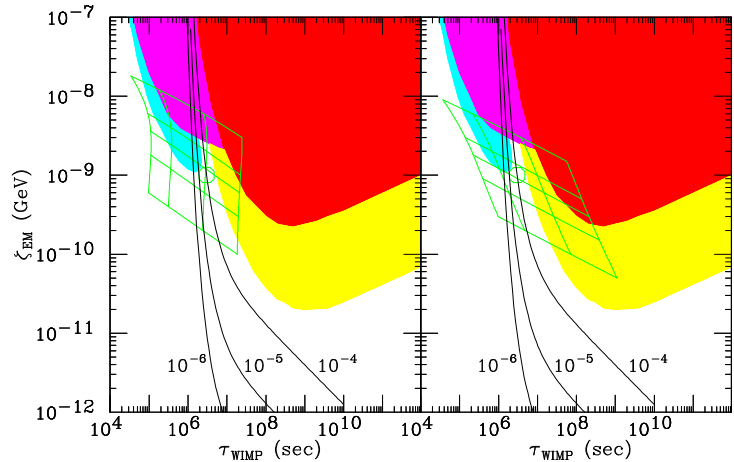}{0.6}
\end{center}
\vspace*{-.42in}
\caption{The grid gives predictions in the superWIMP gravitino dark
  matter scenario for decay time $\tau_{\text{WIMP}}$ and energy
  release $\zeta_{\text{EM}}$ from $\tilde{\gamma} \to \gamma
  \tilde{G}$ (left) and $\tilde{l} \to l \tilde{G}$ (right).  The
  shaded regions are excluded by BBN~\protect\cite{Cyburt:2002uv}, and
  the contours give values for $\mu$ distortions of the
  CMB~\protect\cite{Feng:2003uy}.}
\label{fig:mu}
\end{figure*}

The BBN-excluded regions are shaded in \figref{mu}, and the predicted
values of the superWIMP scenario are given by the grid.  We see that
BBN excludes some of the parameter space, but leaves much of it
intact.  In fact, inconsistencies in the standard BBN picture, notably
the prediction of more $^7$Li than is observed, currently prefer the
decay time and energy release indicated by the circles in \figref{mu}.
SuperWIMP gravitino dark matter may not only pass BBN constraints, but
may even resolve the leading current anomaly in standard BBN.

Can such a scenario be verified?  In fact, it can --- energy release
in the early universe may also leave its imprint on the cosmic
microwave background (CMB).  Parameterizing the CMB spectral shape by
\begin{equation}
f(E) = \frac{1}{e^{E/(kT) + \mu} - 1} \ ,
\end{equation}
electromagnetic energy released at time $\tau \sim \ \text{year}$
cannot be completely thermalized, leading to $\mu > 0$.  The predicted
size of such ``$\mu$ distortions'' from decays to superWIMPs is given
in \figref{mu}.  At present, the CMB is consistent with a Planckian
spectrum; the current bound is $\mu < 9 \times 10^{-5}$.  However, the
Diffuse Microwave Emission Survey (DIMES), a future space mission, is
designed to probe $\mu$ distortions as low as $\mu \sim 10^{-6}$ and
may discover $\mu$ distortions predicted in the superWIMP dark matter
scenario.

\begin{figure}[tbh]
\vspace{9pt}
\postscript{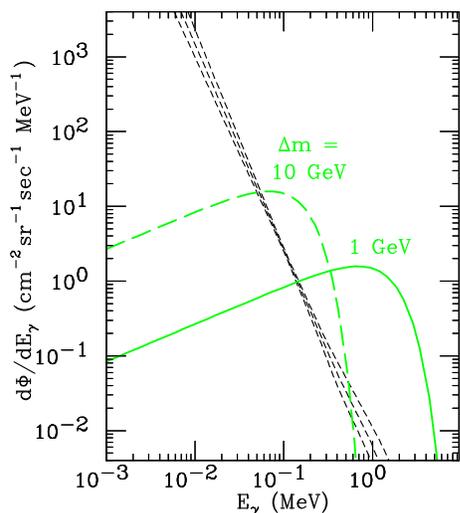}{0.8}
\vspace*{-.42in}
\caption{Diffuse photon flux predictions from decays $\tilde{\gamma}
  \to \gamma \tilde{G}$ in the superWIMP gravitino dark matter
  scenario~\protect\cite{Feng:2003xh}.  The straight lines are
  existing measurements from HEAO, OSSE, and COMPTEL.}
\label{fig:diffuseflux}
\end{figure}

If WIMPs and gravitinos are highly degenerate, WIMP decay may be very
late.  In the case of $\tilde{\gamma} \to \gamma \tilde{G}$, the
produced photons may then be observed as bumps in the diffuse photon
background.  Predicted spectra are shown in \figref{diffuseflux}.  For
the parameters indicated, the flux excesses are already excluded, but
for other underlying supersymmetry parameters, the fluxes may be
reduced to within current uncertainties.  The robust prediction,
however, is that any excess must occur in the keV to MeV range.  Such
signals may again by uncovered by space-based experiments, such as
INTEGRAL now underway. Other implications of late decays have recently
been considered in Refs.~\cite{Chen:2003gz,Sigurdson:2003vy,%
Ellis:2003dn,Buchmuller:2004rq,Jedamzik:2004er,Kawasaki:2004yh,%
Feng:2004zu,Feng:2004mt}.

\section{Summary and Outlook}

We have reviewed some recent advances and their implications in the
study of neutralino dark matter, Kaluza-Klein dark matter, and
superWIMP gravitino dark matter.  Space-based experiments, ranging
from PAMELA and AMS, to GLAST, DIMES, INTEGRAL, and many others, may
shed light on these possibilities and, in some cases, may provide
information that is impossible or very difficult to obtain in
terrestrial experiments.  These dark matter candidates satisfy the
Martha Stewart criterion.  They are therefore especially interesting
as they naturally explain the observed dark matter relic density.  Of
course, there are many other known dark matter candidates and likely
also many more to be discovered. Space-based experiments provide a
window on many of these other possibilities also.

\begin{figure*}[tbh]
\vspace{9pt}
\begin{center}
\postscript{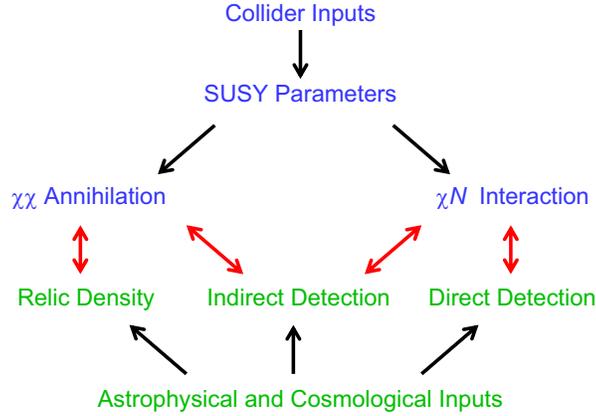}{0.49}
\end{center}
\vspace*{-.42in}
\caption{A schematic picture of the investigation of neutralino dark
  matter from the combined approaches of particle physics,
  astrophysics, and cosmology.}
\label{fig:flowchart}
\end{figure*}

At the same time, although the importance of space-based experiments
has been emphasized here, it is clear that the solution to the dark
matter problem, even in the most favorable of circumstances, will
require a multi-faceted approach drawing on inputs from particle
physics, astrophysics, and cosmology.  A schematic picture of how the
different pieces might fit together for the case of neutralino dark
matter is shown in \figref{flowchart}.  By producing supersymmetric
particles in the laboratory, the microscopic properties of neutralinos
and other supersymmetric particles will be determined.  These will
then determine dark matter properties, such as annihilation rates and
interaction cross sections.  In parallel, astrophysical experiments
and cosmological observations will be able to determine the relic
density with even greater precision, and also possibly detect dark
matter in a variety of ways.  Only by combining all of these
approaches can we hope to develop a compelling microscopic description
of particle dark matter in the coming years.

\vspace*{.1in}

{\em Acknowledgments} --- I thank the organizers of SpacePart '03 for
the invitation to take part in a stimulating and beautifully organized
conference.  This work was supported in part by National Science
Foundation CAREER Award PHY--0239817 and in part by the Alfred
P.~Sloan Foundation.

\end{document}